\documentclass[%
 reprint,
 amsmath,amssymb,
 aps,
]{revtex4-2}

\usepackage{sidecap}
\usepackage{array}
\usepackage{subfigure}
\usepackage{color}
\usepackage{lineno}

\usepackage{hyperref}
\usepackage[dvipsnames]{xcolor}
\usepackage{amsmath}
\usepackage{graphicx}% Include figure files
\usepackage{dcolumn}% Align table columns on decimal point
\usepackage{bm}% bold math

\begin{document}

\title{The elliptic flow difference between baryons and anti-baryons in heavy collision with SMASH Model\thanks{This is a footnote for the title}}
\thanks{This work is supported in part by the National Key Research and Development Program of China under Contract No. 2022YFA1604900 and 2024YFA1610700; the National Natural Science Foundation of China (NSFC) under contract No. 12175084.
The numerical calculation have been performed on the GPU cluster in the Nuclear Science Computing Center at Central China Normal University (NSC3).
}%

\author{Shuai Zhou}

\affiliation{Key Laboratory of Quark and Lepton Physics (MOE) and Institute of Particle Physics, Central China Normal University, Wuhan 430079, China}

\author{Shusu Shi}

\email{shiss@ccnu.edu.cn (Corresponding author)}
\affiliation{Key Laboratory of Quark and Lepton Physics (MOE) and Institute of Particle Physics, Central China Normal University, Wuhan 430079, China}

\date{\today}

%----------------------------------------------------------------------------------------------------------------------------
\begin{abstract}

\textbf{Abstract}: A significant difference in the elliptic flow $v_2$ for particles and their corresponding antiparticles, which is more pronounced for baryons and anti-baryons, was observed in the STAR experiment during the Beam Energy Scan I (BES-I) at RHIC.
By employing the SMASH model, we study the $v_2$ difference between protons and anti-protons, as well as between $\Lambda$ and $\bar{\Lambda}$, in Au + Au collisions at $\sqrt{s_{NN}} = 7.7$ GeV as a function of evolution time.
It finds that as the evolution time increases, the $v_2$ difference between protons and anti-protons becomes more pronounced compared to that between $\Lambda$ and $\bar{\Lambda}$.
This phenomenon can be explained by the different constituent quarks of protons and $\Lambda$. Since some of the $u$ quarks and $d$ quarks come from the colliding nuclei and are transported to mid-rapidity, they undergo more interactions than produced quarks, resulting in the proton $\Delta v_2$ being larger than the $\Lambda$ $\Delta v_2$. We compare the SMASH calculations with STAR BES-I data and argue that higher precision data from BES-II will provide constraints on theoretical frameworks to interpret the $v_2$ differences of particles and anti-particles.


\end{abstract}
%----------------------------------------------------------------------------------------------------------------------------
%\authorrunning
%\titlerunning
\maketitle
\section*{I. Introduction}

During high-energy heavy-ion collisions, the system created reaches extreme conditions of temperature and density, facilitating the study of strongly interacting matter. One of the most significant phenomena observed in such collisions is collective flow ($v_n$), which refers to the correlated motion of particles produced in the collision. Collective flow provides essential information about the early-stage dynamics of the system, the equation of state, and the transport properties of the medium. In heavy-ion collisions, the produced particles do not scatter isotropically, as would be expected from a purely thermal system. Instead, they exhibit a pattern of anisotropic distribution in momentum space, driven by pressure gradients generated in the initial stages of the collision. This anisotropy, commonly decomposed into harmonic coefficients, offers insights into the medium's response to initial geometric asymmetries\cite{1, 2}. Collective flow, thus, plays an important role in understanding the evolution of the system, connecting the microscopic interactions to the macroscopic observable effects. It not only provides a window into the properties of the strongly interacting matter but also serves as a key observable in constraining theoretical models of the collision dynamics\cite{3,4,5,6,7,8,9,10,11,12,13,14,15,16,17,18,19,20}.

 Elliptic flow, denoted as $v_2$, measures the anisotropic distribution of particles produced in the azimuthal direction with respect to the reaction plane ($\Psi_{rp}$) in heavy-ion collisions. It is defined as the second Fourier coefficient in the Fourier decomposition of the particle yield as a function of the azimuthal angle \cite{1}:
\\
\begin{equation} \label{eq1}
%\begin{split}
E\frac{d^{3}N }{d^{3}p} = \frac{1}{2\pi }\frac{d^{2}N }{p_{T}dp_{T}dy  }(1+\sum_{n=1}^{\infty }2v_{_{n} }\cos \left [ n\left ( \phi-\Psi_{rp}   \right )  \right ])
%\frac{1}{2\pi }\frac{d^{2}N }{p_{T}dp_{T}dy  } (1+\sum_{n=1}^{\infty }2v_{_{n} }\cos \left [ n\left ( \phi-\Psi_{rp}   \right )  \right ]).
%\end{split}
\end{equation}

where $\phi$ is the azimuthal angle of the final state particles in the laboratory frame.
The quantity $v_2$ is driven by the initial geometric anisotropy. Specifically, in non-central collisions, where the overlap region is elliptical rather than circular, the anisotropies in particle momentum distributions originate from the initial asymmetries in the system's geometry, leading to the formation of $v_2$.
$v_2$ is a valuable probe for studying the degrees of freedom of nuclear matter created in heavy-ion collisions\cite{21,22,23,24}. The degrees of freedom, whether dominated by partonic or hadronic interactions, are crucial for understanding the QCD phase transition\cite{25,26}.
In addition, the energy dependence of the in-medium correction factor for the in-medium nucleon-nucleon elastic cross section (NNECS) has a significant impact on collective flows~\cite{27}.
Comparing experimental $v_2$ data with theoretical calculations provides information about the degrees of freedom achieved in heavy-ion collisions and further constrains the Equation of State (EoS) of nuclear matter~\cite{27,28,29,30,31,32,33,34}.

The STAR collaboration has measured $v_2$ for identified particles at midrapidity in Au + Au collisions as part of the Beam Energy Scan program at the Relativistic Heavy Ion Collider (RHIC), with collision energies ranging from 7.7 to 62.4 GeV~\cite{35,36,37,38}. A collision energy-dependent difference in $v_2$ between particles and their corresponding antiparticles was observed. This difference increases with decreasing beam energy and is more pronounced for baryons compared to mesons. These results suggest that, at lower energies, particles and antiparticles do not exhibit the universal number-of-constituent-quark scaling of $v_2$ that was observed at 200 GeV~\cite{9,10,12,39}.

Some models have explained the $v_2$ difference from different perspectives. Within the framework of the UrQMD model, tracing the number of initial quarks in protons partly accounts for the difference in $v_2$ between transported protons and anti-protons, providing an explanation for the observed difference in the Beam Energy Scan program at RHIC \cite{40}. In a hybrid model, the fluid dynamical evolution of the fireball is combined with a transport treatment for both the initial state and the final hadronic phase. It has been found that the propagation of the baryon-number current during hydrodynamic evolution, as well as the transport treatment of the hadronic phase, are essential for reproducing the experimental data \cite{41}.
In the context of the quark coalescence model, it is noted that due to baryon stopping, especially at lower energies, different constituent quarks exhibit varying $v_2$. Additionally, differing transport properties lead to distinct interactions among quarks, resulting in different $v_2$ values \cite{42}. The study of the $v_2$ difference remains an area of active research \cite{43,44,45,46,47}. However, due to the limitations imposed by the statistical constraints of the data, we are unable to draw clear conclusions regarding the $v_2$ difference among different baryon types, particularly the relationship between the $v_2$ differences of protons and $\Lambda$ , based on the STAR Beam Scan I results \cite{35,36,37,38}.

In this work, we focus on the relationship between the $v_2$ difference of protons and $\Lambda$ using the Simulating Many Accelerated Strongly-interacting Hadrons (SMASH) model. Since the $v_2$ difference is significant at 7.7 GeV, which is the lowest collision energy in RHIC-STAR Au+Au collisions, we choose this energy for our study. Our goal is to investigate the $v_2$ difference of protons and $\Lambda$ for different centralities and at different times during the collision evolution, specifically at 2 fm/$c$, 10 fm/$c$, and 50 fm/$c$ after the collision. By comparing the $v_2$ difference of protons and $\Lambda$  at these specific times during the collision evolution, we aim to elucidate the dependence of the $v_2$ difference on both transported and produced quarks, as well as on quark flavor in baryons. Additionally, we propose a method to re-examine the mechanisms that explain the $v_2$ difference across different models.

The rest of this paper is organized as follows. We introduce the SMASH model in Sec. II. In Sec. III, we discuss the relationship between the $v_2$ and the collision evolution time, along with its physical interpretation. We provide a summary in Sec. IV.

\section*{II. SMASH MODEL}

\begin{figure*}[]
     \centering
     \resizebox{17.5cm}{!}{%
    \includegraphics[bb = 0.833555 4.002000 600.879960 220.127971,clip]{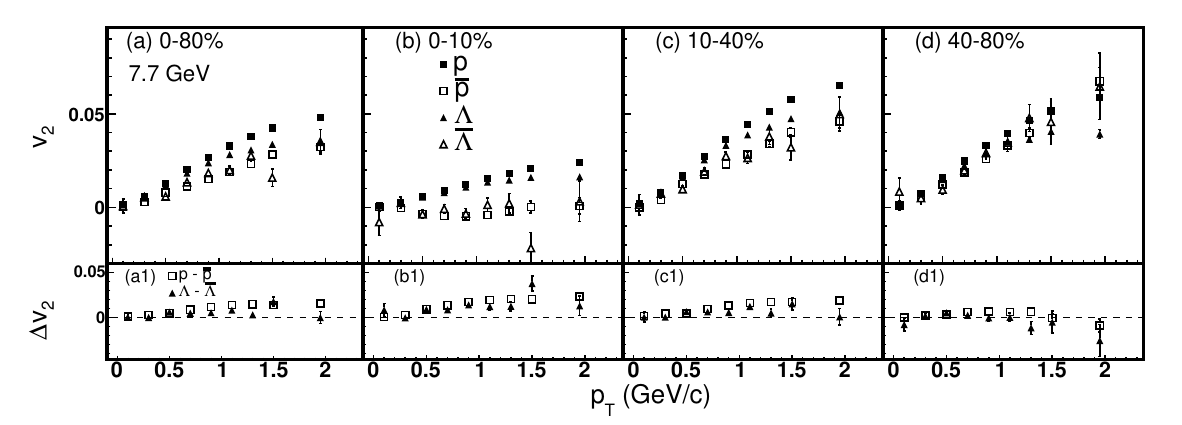}}
    \caption{The $v_2$ as a function of transverse momentum ($p_T$) for protons, antiprotons, $\Lambda$ and $\bar{\Lambda}$ in Au+Au collisions at  \(\sqrt{s_{\text{NN}}} = 7.7 \text{ GeV}\) for four centrality bins: (a) 0-80$\%$, (b) 0-10$\%$, (c) 10-40$\%$, and (d) 40-80$\%$, with a fixed evolution time of 10 fm/$c$. The bottom panels display the $v_2$ differences of baryons and anti-baryons as a function of $p_{T}$ accordingly.}
     \label{fv2_proton_pt_diff_10fm}
\end{figure*}

\begin{figure*}[]
    \centering
        \resizebox{17.5cm}{!}{%
        \includegraphics[bb = 0.833555 3.002000 600.879960 220.127971,clip]{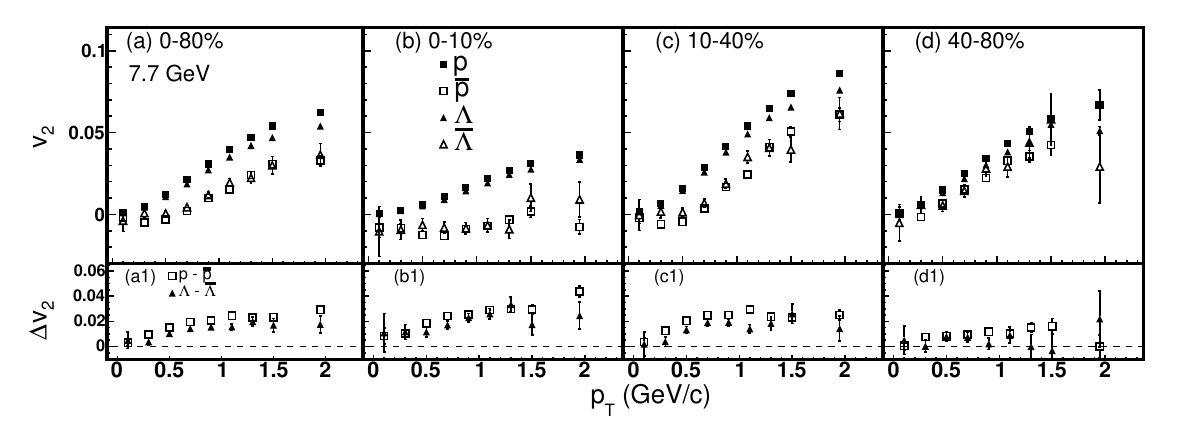}}
        \caption{The $v_2$ as a function of $p_T$ for protons, antiprotons, $\Lambda$ and $\bar{\Lambda}$ in Au+Au collisions at  \(\sqrt{s_{\text{NN}}} = 7.7 \text{ GeV}\) for four centrality bins: (a) 0-80$\%$, (b) 0-10$\%$, (c) 10-40$\%$, and (d) 40-80$\%$, with a fixed evolution time of 50 fm/$c$. The bottom panels display the $v_2$ differences of baryons and anti-baryons as a function of $p_{T}$ accordingly.}
        \label{fv2_Lambda_pt_diff_50fm}
\end{figure*}

\begin{figure*}[]
  \centering
  \resizebox{17.5cm}{!}{%
    \includegraphics[bb = 0.833555 5.002000 600.879960 220.127971,clip]{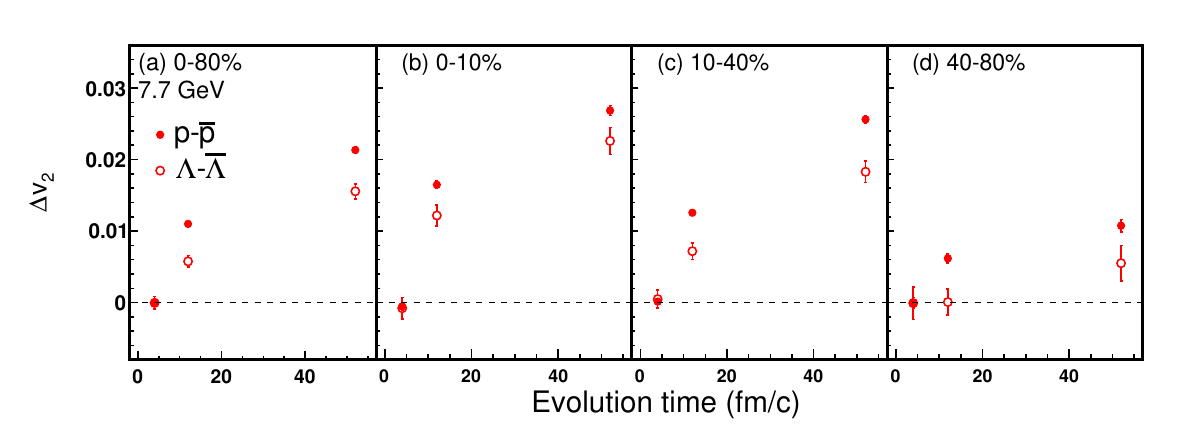}}
    \caption{The $v_{2}$ differences of baryons and anti-baryons as a function of evolution time in Au+Au collision at \(\sqrt{s_{\text{NN}}} = 7.7 \text{ GeV}\) for four centrality bins: (a) 0-80$\%$, (b) 0-10$\%$, (c) 10-40$\%$, and (d) 40-80$\%$.}
  \label{fv2_proton_pt_diff}
  \end{figure*}

To model the dynamics of heavy-ion collisions, various theoretical frameworks have been developed. Among these, the SMASH model is a state-of-the-art hadronic transport approach that simulates the late-stage hadronic interactions following the partonic stage \cite{48,49}. 
The research advances within the SMASH framework highlight significant developments in understanding the dynamics of heavy-ion collisions. Specifically, SMASH has been instrumental in elucidating the mechanisms of light nuclei formation, baryon stopping, and the generation of anisotropies in momentum space. By integrating advanced models such as CRAB for coalescence studies\cite{50} and employing sophisticated parameterizations of the shear viscosity $\eta$/s\cite{51}, SMASH has successfully reproduced experimental observations of particle correlations and flow patterns\cite{52}. These advancements underscore the capability of model to accurately describe the complex physics of heavy-ion collisions across a wide range of energies.
Additionally, they contribute to a deeper understanding of the QCD phase diagram and the properties of nuclear matter under extreme conditions.
Future enhancements will focus on refining the initial conditions and improving the parametrization of the EoS, further solidifying the role of SMASH in advancing our knowledge of strong interaction dynamics\cite{53}.

SMASH incorporates a comprehensive treatment of hadronic interactions, including binary collisions, resonance decays, and the propagation of hadrons through the medium. It is particularly suitable for low-energy heavy-ion collisions and provides a detailed description of the freeze-out process and the development of collective flow\cite{54}. For the specific scenario of 7.7 GeV Au + Au collisions, we have selected time points at 2 fm/$c$, 10 fm/$c$, and 50 fm/$c$ as the endpoints of the cascade mode simulation. In the context of system evolution, the early stage is characterized by a timescale of approximately 2 fm/$c$. The intermediate to late stages are associated with a duration of around 10 fm/$c$, while the final stage of the system's evolution is defined by a timescale of approximately 50 fm/$c$. These time points represent key stages in the collision's evolution, allowing us to study the dynamics of particle interactions and the development of collective flow phenomena\cite{55}. Employing a consistent time step size and focusing on these specific time points, we aim to obtain a comprehensive understanding of the evolution of the system while maintaining computational efficiency in our analysis.
We calculated the elliptic flow using $v_{2} = \cos \left( 2\phi \right)$, since the azimuth of the reaction plane is set to 0 in the model. The event centrality is defined based on the collision impact parameter, because in the early stages of the collision, the production of particles is not yet fully formed. Therefore, it is not appropriate to define centrality using the reference multiplicity of final-state particles.

\begin{figure*}[]
  \centering
  \resizebox{12cm}{!}{%
    \includegraphics[bb = 0.833555 5.002000 600.879960 420.127971,clip]{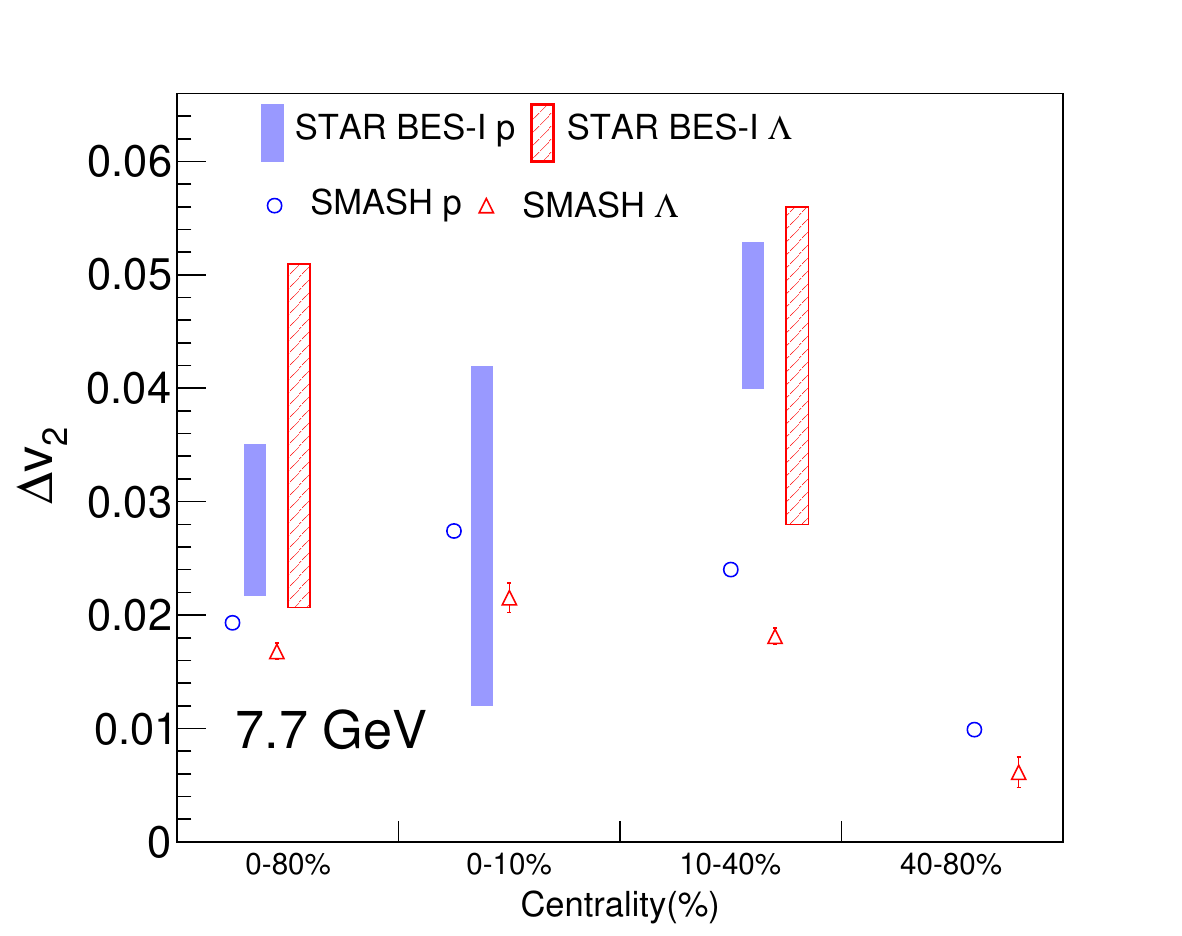}}
    \caption{The $v_2$ differences between baryons and anti-baryons in Au+Au collisions at $\sqrt{s_{NN}} = 7.7 \text{ GeV}$ are presented for four centrality bins: 0-80\%, 0-10\%, 10-40\%, and 40-80\%, using the full default evolution time (200 fm/$c$) in the SMASH model. The model results are compared with the experimental data from STAR BES-I.
}
  \label{fv2_compare_BESI}
  \end{figure*}
  
\section*{III. Results and discussion}

The duration of the collision evolution is referred to as the evolution time. We calculated the $v_{2}$ differences by subtracting the $v_{2}$ of particles from that of their corresponding antiparticles within each transverse momentum ($p_T$) bin. Subsequently, a fitting procedure was employed to determine the mean $v_{2}$ difference by analyzing the individual $v_{2}$ differences over the $p_T$ range from 0.6 to 2.3 GeV/$c$, similar to the method used in the STAR experiment~\cite{35,36,37}. This approach allows for the characterization of the $v_{2}$ difference between particles and antiparticles across a range of momenta, providing insights into the asymmetries in the azimuthal anisotropy of the final-state particles.

Figure~\ref{fv2_proton_pt_diff_10fm} presents the proton $v_{2}$ differences and the $\Lambda$ $v_{2}$ differences at an evolution time of 10 fm/$c$ in Au+Au collisions at \(\sqrt{s_{\text{NN}}} = 7.7 \text{ GeV}\). The upper panels (a), (b), (c), and (d) of the figure display the $v_{2}$ for protons, \(\Lambda\), and their corresponding antiparticles. Protons and antiprotons are indicated by squares, while \(\Lambda\) and \(\bar{\Lambda}\) are denoted by triangles. The lower panels (a1), (b1), (c1), and (d1) illustrate the proton $v_{2}$ difference and the $\Lambda$ $v_{2}$ difference for each $p_{T}$ bin. Our findings reveal that across different centrality bins, the $v_{2}$ differences for protons are generally greater than those for \(\Lambda\). At an evolution time of 10 fm/$c$, the majority of particles have undergone freeze-out, with interactions among them significantly diminishing, leading to the stabilization of $v_{2}$ differences. By examining the $v_{2}$ differences between protons and $\Lambda$ particles at this stage of evolution, we aimed to investigate the influence of differing quark compositions on the observed $v_{2}$ variations.

At an evolution time of 50 fm/$c$, compared to 10 fm/$c$, the freeze-out of particles is more complete and the interactions are more sufficient. As shown in panels (a), (b), (c), and (d) of Fig.~\ref{fv2_Lambda_pt_diff_50fm}, the $v_{2}$ magnitude is larger than the results at an evolution time of 10 fm/$c$. The same conclusion applies to the $v_2$ differences in panels (a1), (b1), (c1), and (d1) of Fig. 2. Consequently, choosing 50 fm/$c$ provides a clearer and more definitive illustration of the influence of different quark compositions on the proton $v_{2}$ differences and the $\Lambda$ $v_{2}$ differences. Across different centrality intervals, the proton $v_{2}$ differences are markedly greater than the $\Lambda$ $v_{2}$ differences. This observation further strengthens the evidence for the critical role of varying quark configurations in the $v_{2}$ difference observed in heavy-ion collisions.

In Fig.~\ref{fv2_proton_pt_diff}, the extracted results of the proton $v_{2}$ difference and the $\Lambda$ $v_{2}$ difference as functions of evolution time are shown. 
It is evident that at 2 fm/$c$, the proton $v_{2}$ difference closely resembles the $\Lambda$ $v_{2}$ difference, both converging near zero for each centrality bin.
As time progresses, the magnitude of $v_2$ increases, and the distinction between the proton $v_{2}$ difference and the $\Lambda$ $v_{2}$ difference becomes increasingly pronounced.
Notably, at 10 fm/$c$ and 50 fm/$c$, the proton $v_{2}$ difference exceeds the $\Lambda$ $v_{2}$ difference, highlighting a growing disparity between the two quantities.
As previously discussed, an evolution time of 2 fm/$c$ corresponds to the early stages of the collision.  During this early stage, interactions among particles are rare, and only a few particles are produced. As previously stated, $v_{2}$ depends on these interactions. Consequently, at an evolution time of 2 fm/$c$, the $v_{2}$ differences are nearly zero for the four centrality bins. By examining the $v_{2}$ differences during the early stages of the collision and at early evolution times, we can gain a more comprehensive understanding of the interactions between particles and their impact on the $v_{2}$ of protons and \(\Lambda\). These calculations provide insights into the fundamental dynamics governing the early-time evolution of the system and the subsequent development of $v_{2}$.

At an evolution time of 10 fm/$c$, both the proton $v_{2}$ and the $\Lambda$ $v_{2}$ increase significantly. 
At the same time, the $v_{2}$ of baryons and anti-baryons shows clear differences and increases as the evolution time increases.  
Considering the composition of different constituent quarks in the anti-proton and $\overline{\Lambda}$, with the proton containing three constituent quarks ($u$$u$$d$), the anti-proton consisting of $\overline{u}$, $\overline{u}$, $\overline{d}$ constituent quarks, the $\Lambda$ consisting of $u$, $d$, $s$ constituent quarks, and $\overline{\Lambda}$ consisting of $\overline{u}$, $\overline{d}$, $\overline{s}$ constituent quarks, we can delve into the dynamics within the constituent quark paradigm.
Particles that are transported from the beam rapidity (\(y = y_{\text{beam}}\)) to midrapidity (\(y = 0\)) undergo a greater number of interactions, whereas particles and antiparticles produced from the vacuum at lower energies are likely to experience fewer interactions. 
In other words, $u$ quarks and $d$ quarks, which are present from the onset of the collision and experience the transport process, undergo more interactions compared to $s$, $\overline{s}$, $\overline{u}$,  $\overline{d}$ quarks. 
Under this scenario, one can infer that transported particles exhibit a larger $v_2$.
Considering that $s$ quarks are also produced, they may encounter fewer collisions, leading to lower $v_{2}$ values. Therefore, we infer that the $\Lambda$ ($uds$) $v_{2}$ is lower than the proton ($uud$) $v_{2}$. 
Similarly, given that $\overline{u}$, $\overline{d}$, and $\overline{s}$ are produced quarks, the $v_2$ value for $\overline{\Lambda}$ ($\overline{u}$$\overline{d}$$\overline{s}$) is comparable to that for the anti-proton ($\overline{u}$$\overline{u}$$\overline{d}$). The $v_2$ value for the anti-proton ($\overline{u}$$\overline{u}$$\overline{d}$) is lower than that for the proton ($uud$), and the $v_2$ value for the $\overline{\Lambda}$ ($\overline{u}$$\overline{d}$$\overline{s}$) is less than that for the $\Lambda$ ($uds$).

We finally compare the SMASH calculations with the existing STAR BES-I experimental data.
Figure~\ref{fv2_compare_BESI} illustrates the proton $v_{2}$ differences and $\Lambda$ $v_{2}$ differences for different centrality bins in Au + Au collisions at \(\sqrt{s_{\text{NN}}} = 7.7 \text{ GeV}\). The blue bands represent the $v_{2}$ differences of protons and anti-protons from STAR BES-I in the 0-80\%, 0-10\% and 10-40\% centrality bins, while red band represents the $v_{2}$ difference of $\Lambda$ and $\bar{\Lambda}$ from STAR BES-I in 0-80\% centrality range~\cite{35,36,37}. The open blue circles and open red triangles correspond to the results from the SMASH model for proton and $\Lambda$ $v_{2}$ differences, respectively.
The fitting ranges used to extract the $v_{2}$ differences in the SMASH model are consistent with the STAR BES-I experimental data for each centrality bin. 
For protons in the centrality bins of 0-80\%, 0-10\%, and 10-40\%, the $p_T$ fitting ranges used are 0.46-1.64 GeV/$c$, 0.78-1.49 GeV/$c$, and 0.46-1.64 GeV/$c$ respectively, consistent with STAR BES-I.  
For $\Lambda$ in the 0-80\% and 10-40\% centrality bins, we use the same $p_T$ fitting ranges, 0.87-1.53 GeV/$c$ and 0.46-1.64 GeV/$c$, as those in STAR BES-I. 
In the 0-10\% centrality bin, since no experimental data is available for $\Lambda$ $\Delta v_2$, we apply the same fitting range as that used for protons to ensure a fair comparison.
For the 40-80\% centrality bin, with no STAR BES-I results to reference, we use the fitting range of 0.60-2.3 GeV/$c$, following the approach in Fig.~\ref{fv2_proton_pt_diff}.
The default evolution time in the SMASH model is 200 fm/$c$.
As shown in Fig.~\ref{fv2_compare_BESI}, we observe that the SMASH model quantitatively describes the STAR BES-I experimental data in the 0-10\% centrality bins within the large data uncertainties. Due to the large statistical errors in the BES-I data, it is not possible to conclude whether the model results quantitatively reproduce the data in the 0-80\% and 10-40\% centrality bins. Higher statistics from BES-II are needed to address this.
Given the relatively low statistics and larger uncertainties in the STAR BES-I data, it is not possible to definitively determine whether the $\Delta v_2$ differs between protons and $\Lambda$ particles. Furthermore, research suggests that the effects induced by a strong electromagnetic field may influence the observed $v_{2}$ differences\cite{56,57}.

Most importantly, it is noted that the proton $v_{2}$ differences are greater than the $\Lambda$ $v_{2}$ differences in the SMASH model. The baryon-type dependence of the $v_{2}$ difference between baryons and anti-baryons can be explained by the transported and produced particles discussed above. The BES-I data cannot verify or falsify this, as the uncertainties are large. The new measurements from STAR BES-II will provide experimental data with higher precision and further constrain the models\cite{40,41,42,43,44,45,46,47}.

\section*{IV. Summary}

In this paper, using the SMASH model, we analyze the $v_{2}$ of protons and $\Lambda$ in Au + Au collisions at $\sqrt{s_{NN}}$ = 7.7 GeV as a function of evolution time. 
The results show that the $v_{2}$ differences are nearly zero at 2 fm/$c$ for protons ($\Lambda$) and their corresponding antiparticles. However, as the evolution time progresses, the disparity between the proton $v_{2}$ difference and the $\Lambda$ $v_{2}$ difference becomes increasingly prominent, with the proton $v_{2}$ difference surpassing that of the $\Lambda$.
The explanation is that protons contain more transported quarks compared to $\Lambda$ from a statistical perspective. These transported quarks, having undergone more interactions, exhibit larger $v_{2}$ values. As the evolution time increases, a more pronounced difference develops between the $\Delta v_2$ of protons and that of $\Lambda$.
We compared the $\Delta v_2$ results of protons and $\Lambda$ with the data from the STAR BES-I experiment and observed that the SMASH results quantitatively match the experimental data in the 0-80\% and 0-10\% centrality bins.

Due to the statistical limitations of the experimental data, one cannot distinguish whether there is a difference between proton and $\Lambda$ $\Delta v_2$.  
The STAR BES-II experiment recorded 12-25 times more events than BES-I for Au + Au collisions at  $\sqrt{s_{NN}}$ = 7.7 - 19.6 GeV. 
The incoming new measurements from STAR BES-II will provide constraints on the model interpretations for the $v_2$ differences between particles and antiparticles.

% \begin{acknowledgments}
% This work is supported in part by the National Key Research and Development Program of China under Contract No. 2022YFA1604900; the National Natural Science Foundation of China (NSFC) under contract No. 12175084.
% \end{acknowledgments}

%\printbibliography


\begin{thebibliography}{0}%
\makeatletter
\providecommand \@ifxundefined [1]{%
 \@ifx{#1\undefined}
}%
\providecommand \@ifnum [1]{%
 \ifnum #1\expandafter \@firstoftwo
 \else \expandafter \@secondoftwo
 \fi
}%
\providecommand \@ifx [1]{%
 \ifx #1\expandafter \@firstoftwo
 \else \expandafter \@secondoftwo
 \fi
}%
\providecommand \natexlab [1]{#1}%
\providecommand \enquote  [1]{``#1''}%
\providecommand \bibnamefont  [1]{#1}%
\providecommand \bibfnamefont [1]{#1}%
\providecommand \citenamefont [1]{#1}%
\providecommand \href@noop [0]{\@secondoftwo}%
\providecommand \href [0]{\begingroup \@sanitize@url \@href}%
\providecommand \@href[1]{\@@startlink{#1}\@@href}%
\providecommand \@@href[1]{\endgroup#1\@@endlink}%
\providecommand \@sanitize@url [0]{\catcode `\\12\catcode `\$12\catcode
  `\&12\catcode `\#12\catcode `\^12\catcode `\_12\catcode `\%12\relax}%
\providecommand \@@startlink[1]{}%
\providecommand \@@endlink[0]{}%
\providecommand \url  [0]{\begingroup\@sanitize@url \@url }%
\providecommand \@url [1]{\endgroup\@href {#1}{\urlprefix }}%
\providecommand \urlprefix  [0]{URL }%
\providecommand \Eprint [0]{\href }%
\providecommand \doibase [0]{https://doi.org/}%
\providecommand \selectlanguage [0]{\@gobble}%
\providecommand \bibinfo  [0]{\@secondoftwo}%
\providecommand \bibfield  [0]{\@secondoftwo}%
\providecommand \translation [1]{[#1]}%
\providecommand \BibitemOpen [0]{}%
\providecommand \bibitemStop [0]{}%
\providecommand \bibitemNoStop [0]{.\EOS\space}%
\providecommand \EOS [0]{\spacefactor3000\relax}%
\providecommand \BibitemShut  [1]{\csname bibitem#1\endcsname}%
\let\auto@bib@innerbib\@empty
%</preamble>
\end{thebibliography}%


\begin{thebibliography}{99}\footnotesize
\itemsep=-1pt plus.2pt minus.2pt

\bibitem {1} Voloshin S A and Zhang Y 1996 {\it Z. Phys. C} {\bf 70} 665-672

\bibitem {2} Poskanzer A M and Voloshin S A 1998 {\it Phys. Rev. C} {\bf 58} 1671-1678

\bibitem {3} Voloshin S A, Poskanzer A M and Snellings R 2010 {\it Landolt-Bornstein} {\bf 23} 293-333

\bibitem {4} Bzdak A, Esumi S, Koch V, Liao J F, Stephanov M and Xu N 2020 {\it Phys. Rept.} {\bf 853} 1-87

\bibitem {5} Luo X F, Shi S S, Xu N and Zhang Y F 2020 {\it Particles} {\bf 3} 278-307

\bibitem {6} Chen J H {\it et al}. 2024 {\it Nucl. Sci. Tech.} {\bf 12} 214

\bibitem {7} Abdallah M S {\it et al}. (STAR Collaboration) 2022 {\it Phys. Lett. B} {\bf 827} 136941

\bibitem {8} Abdallah M S  {\it et al}. (STAR Collaboration) 2022 {\it Phys. Lett. B} {\bf 827} 137003

\bibitem {9} Adamczyk L {\it et al}. (STAR Collaboration) 2017 {\it Phys. Rev. Lett.} {\bf 118} 212301

\bibitem {10} Adamczyk L {\it et al}. (STAR Collaboration) 2016 {\it Phys. Rev. Lett.} {\bf 116} 062301

\bibitem {11} Adamczyk L {\it et al}. (STAR Collaboration) 2015 {\it Phys. Rev. Lett.} {\bf 115} 222301

\bibitem {12} Abelev B I {\it et al}. (STAR Collaboration) 2010 {\it Phys. Rev. C} {\bf 81} 044902

\bibitem {13} Ollitrault J Y 1992 {\it Phys. Rev. D} {\bf 46} 229-245

\bibitem {14} Gustafsson H A {\it et al}. 1984 {\it Phys. Rev. Lett.} {\bf 52} 1590

\bibitem {15} Aamodt K {\it et al}. (ALICE Collaboration) 2010 {\it Phys. Rev. Lett.} {\bf 105} 252302

\bibitem {16} Adler S S {\it et al}. (PHENIX Collaboration) 2003 {\it Phys. Rev. Lett.} {\bf 91} 182301

\bibitem {17} Chatrchyan S {\it et al}. (CMS Collaboration) 2013 {\it Phys. Rev. C} {\bf 87} 014902

\bibitem {18} Liu Z W and Shi S S 2024 {\it Phys. Rev. C} {\bf 110} 034903

\bibitem {19} Zhao S J, Xu H J, Zhou Y, Liu Y X and Song H C 2024 {\it Phys. Rev. Lett.} {\bf 133} 192301

\bibitem {20} Jia J Y, Giacalone G and Zhang C J 2023 {\it Chin. Phys. Lett.} {\bf 40} 042501

\bibitem {21} Xu Z Y, Liu J L, Zhang P P, Zhang J B and H Lei 2017 {\it Chin. Phys. Lett.} {\bf 34} 062501

\bibitem {22} Yan T Z 2013 {\it Chin. Phys. Lett.} {\bf 30} 092501

\bibitem {23} Yang H Y, Zhou D C, Yang C B and Cai X 2002 {\it Chin. Phys. Lett.} {\bf 19} 1082

\bibitem {24} Liu J L, Shan L Q, Feng Q C, Wu F J, Zhang J B, Tang G X and Huo L 2009 {\it Chin. Phys. Lett.} {\bf 26} 072501

\bibitem {25} Li N and Shi S S 2011 {\it Chin. Phys. Lett.} {\bf 28} 122502

\bibitem {26} Wu S, Shen C and Song H C 2021 {\it Chin. Phys. Lett.} {\bf 38} 081201

\bibitem {27} Li P C, Wang Y J, Li Q F and Zhang H F 2022 {\it Phys. Lett. B} {\bf 828} 137019

\bibitem {28} Colonna M 2020 {\it Prog. Part. Nucl. Phys.} {\bf 113} 103775

\bibitem {29} Huovinen P, Kolb P F, Heinz U W, Ruuskanen P V and Voloshin S A 2001 {\it Phys. Lett. B} {\bf 503} 58

\bibitem {30} Shen C and Heinz U 2012 {\it Phys. Rev. C} {\bf 85} 054902

\bibitem {31} Zhao W, Ko C M, Liu Y X, Qin G Y and Song H 2020 {\it Phys. Rev. Lett.} {\bf 125} 072301

\bibitem {32} Gross D J, Pisarski R D and Yaffe L G 1981 {\it Rev. Mod. Phys.} {\bf 53} 43

\bibitem {33} Lin Z W, Ko C M, Li B A, Zhang B and Pal S 2005 {\it Phys. Rev. C} {\bf 72} 064901

\bibitem {34} Huovinen P and Petreczky P 2010 {\it Nucl. Phys. A} {\bf 837} 26

\bibitem {35} Adamczyk L {\it et al}. (STAR Collaboration) 2016 {\it Phys. Rev. C} {\bf 93} 014907

\bibitem {36} Adamczyk L {\it et al}. (STAR Collaboration) 2013 {\it Phys. Rev. Lett.} {\bf 110} 142301

\bibitem {37} Adamczyk L {\it et al}. (STAR Collaboration) 2013 {\it Phys. Rev. C} {\bf 88} 014902

\bibitem {38} Adamczyk L {\it et al}. (STAR Collaboration) 2012 {\it Phys. Rev. C} {\bf 86} 054908

\bibitem {39} Adams J {\it et al}. (STAR Collaboration) 2005 {\it Phys. Rev. Lett.} {\bf 95} 122301

\bibitem {40} Tu B, Shi S S and Liu F 2019 {\it Chin. Phys. C} {\bf 43} 054106

\bibitem {41} Steinheimer J, Koch V and Bleicher M 2012 {\it Phys. Rev. C} {\bf 86} 044903

\bibitem {42} Dunlop J C, Lisa M A and Sorensen P 2011 {\it Phys. Rev. C} {\bf 84} 044914

\bibitem {43} Xu J, Song T, Ko C M and Li F 2014 {\it Phys. Rev. Lett.} {\bf 112} 012301

\bibitem {44} Hatta Y, Monnai A and Xiao B W 2015 {\it Phys. Rev. D} {\bf 92} 114010

\bibitem {45} Liu H, Wang F T, Sun K J, Xu J and Ko C M 2019 {\it Phys. Lett. B} {\bf 798} 135002

\bibitem {46} Li P C, Wang Y J, Steinheimer J, Li Q F and Zhang H F 2020 {\it Mod. Phys. Lett. A} {\bf 35} 2050289

\bibitem {47} Burnier Y, Kharzeev D E, Liao J F and Yee H U 2011 {\it Phys. Rev. Lett.} {\bf 107} 052303

\bibitem {48} SMASH Collaboration, The SMASH website: {\it https://smash-transport.github.io}

\bibitem {49} Weil J {\it et al}. (SMASH Collaboration) 2016 {\it Phys. Rev. C} {\bf 94} 054905

\bibitem {50} Bailung Y, Rode S P, Shah N and Roy A (2024) arXiv:2407.17962[nucl-th] 

\bibitem {51} Götz N, Constantin L and Elfner H 2023 Workshop for Young Scientists on the Physics of Ultra-relativistic Nucleus-Nucleus Collisions

\bibitem {52} Schäfer A, Karpenko I, Wu X Y, Hammelmann J and Elfner H 2022 {\it Eur. Phys. J. A} {\bf 58} 230

\bibitem {53} Tarasovičová L A, Mohs J, Andronic A, Elfner H and Kampert K H 2024 {\it Eur. Phys. J. A} {\bf 60} 232

\bibitem {54} Petersen H, Oliinychenko D, Mayer M, Staudenmaier J and Ryu S 2019 {\it Nucl. Phys. A} {\bf 982} 399-402

\bibitem {55} Le Fèvre A, Leifels Y, Hartnack C and Aichelin J 2018 {\it Phys. Rev. C} {\bf 98} 034901

\bibitem {56} Acharya S {\it et al}. (ALICE Collaboration) 2020 {\it Phys. Rev. Lett.} {\bf 125} 022301

\bibitem {57} Abdulhamid M I {\it et al}. (STAR Collaboration) 2024 {\it Phys. Rev. X} {\bf 14} 011028

\end{thebibliography}
\end{document}